\documentclass[12pt]{article}
\pdfoutput=1
\usepackage{color}
\usepackage{epsfig, palatino}
\usepackage{epic}
\usepackage{dsfont}
\usepackage{colonequals}
\usepackage{mathrsfs}
\usepackage{ae} 
\usepackage[T1]{fontenc}
\usepackage[ansinew]{inputenc}
\usepackage{amsmath}
\usepackage{amssymb}
\usepackage{graphicx}
\usepackage[normalem]{ulem}
\usepackage{xcolor}
\definecolor{darkblue}{cmyk}{0.9,0.9,0,0}
\usepackage[colorlinks=true,linkcolor=darkblue,citecolor=darkblue,urlcolor=darkblue]{hyperref}
\usepackage{cite}
\usepackage{hyperref}
\usepackage{varioref}
\usepackage{makeidx}
\usepackage[english]{babel}
\usepackage{simplewick}
\usepackage{array}
\usepackage{subcaption}
\usepackage{multirow}
\usepackage{bibentry}

\usepackage[font={small}]{caption}

\newcommand{\comment}[1]{}

\newcommand{\begBvR}[1]{\begin{#1}} 
\newcommand{\beq}{\begBvR{equation}}
\newcommand{\eeq}{\end{equation}}

\newcommand{\eeqq}{\end{equation*}}

\newcommand\eeqaa{\end{eqnarray*}}

\newcommand\eeqa{\end{array}}
\newcommand{\bea}{\begBvR{eqnarray}}
\newcommand{\eea}{\end{eqnarray}}

\newcommand\IM{{\rm Im}\,}
\newcommand\RE{{\rm Re}\,}

\newcommand{\neqa}{\nonumber\end{eqnarray}} 
\newcommand{\la}[1]{\label{#1}}

\newcommand{\p}{\partial}

\renewcommand{\d}{\partial}

\newcommand{\<}{{\langle}}
\renewcommand{\>}{{\rangle}}

\newcommand{\re}{\relax{\rm I\kern-.18em R}}

\renewcommand{\sp}{p\hspace{-.40em}/}

\definecolor{darkgreen}{rgb}{0.0, 0.45, 0.0}
\definecolor{mathematicablue}{RGB}{94,130,182}

\def\XXint#1#2#3{{\setbox0=\hbox{$#1{#2#3}{\int}$}
\vcenter{\hbox{$#2#3$}}\kern-.5\wd0}}

\def\su2{{SU(2)}}

\def\[{\left[}
\def\]{\right]}

\def\({\left(}
\def\){\right)}
\def\[{\left[}
\def\]{\right]}

\def\<{\langle}
\def\>{\rangle}

\def\i2{\frac{i}{2}}

\def\spi{\relax{\rm \pi\kern-0.5em /}}
\def\sA{\relax{\rm A\kern-0.5em /}}
\def\sp{\relax{\rm p\kern-0.5em /}}
\def\sd{\relax{\rm \d\kern-0.5em /}}
\def\sk{\relax{\rm k\kern-0.5em /}}
\def\sn{\relax{\rm n\kern-0.5em /}}
\def\sl{\relax{\rm l\kern-0.5em /}}
\def\sP{\relax{\rm P\kern-0.7em /}}
\def\sBethe{\relax{\rm \Bethe\kern-0.5em /}}

\def\2F1{\,_2{\rm F}_1}

\makeindex

\newcommand\blfootnote[1]{%
  \begingroup
  \renewcommand\thefootnote{}\footnote{\hspace{-6mm}#1}%
  \addtocounter{footnote}{-1}%
  \endgroup
}

\begin{document}

\thispagestyle{empty}

\renewcommand{\thefootnote}{\fnsymbol{footnote}}
\setcounter{page}{1}
\setcounter{footnote}{0}
\setcounter{figure}{0}

\vspace{-0.4in}

\begin{center}
$$$$
{\Large\textbf{\mathversion{bold}
S-matrix Bootstrap for Effective Field Theories: \\
Massless Pions
}\par}
\vspace{1.0cm}

\textrm{Andrea L Guerrieri$^\text{\tiny 1,2}$, Jo\~ao Penedones$^\text{\tiny 3}$, Pedro Vieira$^\text{\tiny 1,\tiny 4}$}
\blfootnote{\tt  \#@gmail.com\&/@\{andrea.leonardo.guerrieri,jpenedones,pedrogvieira\}}
\\ \vspace{1.2cm}
\footnotesize{\textit{
$^\text{\tiny 1}$ICTP South American Institute for Fundamental Research, IFT-UNESP, S\~ao Paulo, SP Brazil 01440-070   \\
$^\text{\tiny 2}$School of Physics and Astronomy, Tel Aviv University, Ramat Aviv 69978, Israel   \\
$^\text{\tiny 3}$Fields and Strings Laboratory, 
Institute of Physics, \'Ecole Polytechnique F\'ed\'erale de Lausanne (EPFL),
 CH-1015 Lausanne,
Switzerland \\
$^\text{\tiny 4}$Perimeter Institute for Theoretical Physics,
Waterloo, Ontario N2L 2Y5, Canada
}  
\vspace{4mm}
}
\end{center}

\par\vspace{1.5cm}


\vspace{2mm}
\begin{abstract}
We use the numerical S-matrix bootstrap method to obtain bounds on the two leading Wilson coefficients (or low energy constants) of the chiral lagrangian controlling the low-energy dynamics of massless pions.
This provides a proof of concept that the numerical S-matrix bootstrap can be used to derive non-perturbative bounds on EFTs in more than two spacetime dimensions.


\end{abstract}

\noindent

\setcounter{page}{1}
\renewcommand{\thefootnote}{\arabic{footnote}}
\setcounter{footnote}{0}

\setcounter{tocdepth}{2}

 \def\nref#1{{(\ref{#1})}}

\newpage

\tableofcontents

\parskip 5pt plus 1pt   \jot = 1.5ex

\newpage
\section{Introduction and main results} \la{intro}

Effective Field Theories (EFT) conveniently describe
gapless systems that are weakly coupled at low energy.
An EFT is characterized by its particle content and a (graded infinite) set of Wilson coefficients controlling the interactions of these particles.
The Wilson coefficients depend on the specific microscopic realization of the system and, generically, are very difficult to compute abinitio. 
For this reason, it is desirable to find universal bounds on these coefficients using only general principles like unitarity and causality \cite{Adams:2006sv, Bellazzini:2016xrt, Vecchi:2007na, Nicolis:2009qm, deRham:2017avq, deRham:2017zjm, NimaCERN, Bellazzini:2020cot,Manohar:2008tc,Wang:2020jxr,Tolley:2020gtv}.

In this article, we apply the numerical S-matrix bootstrap approach to estimate universal bounds on the two leading Wilson coefficients of the EFT describing massless pions, \emph{i.e.} the chiral lagrangian \cite{Weinberg:1978kz, Gasser:1983yg}
\beq
\label{chilag}
\mathcal{L} = \frac{1}{4}f_\pi^2 \,{\rm tr} \left(\partial_\mu U^\dagger \partial^\mu U \right)+
\ell_1 \left[ {\rm tr} \left(\partial_\mu U^\dagger \partial^\mu U \right)\right]^2
+\ell_2\, {\rm tr} \left(\partial_\mu U^\dagger \partial_\nu U \right)  {\rm tr} \left(\partial^\mu U^\dagger \partial^\nu U \right) +\dots
\eeq
where the $SU(2)$ matrix valued field $U(x)=\exp [\frac{i}{f_\pi} \sum_{a=1}^3\sigma^a  \pi^a (x)]$ encodes the pion fields and $f_\pi$ is the pion decay constant.\footnote{We are careless about the distinction between bare and renormalized couplings because we define all parameters from the physical S-matrix \eqref{lowenansatz}.}
The Wilson coefficients $\ell_1$ and $\ell_2$ control the two independent four derivative terms in the effective lagrangian and the dots in \eqref{chilag} denote terms with more than four derivatives.
Using this effective lagrangian, one can compute the low energy behaviour of the four pion scattering amplitude,
\beq
\mathcal{T}_{ab}^{cd} = A(s|t,u) \delta_{ab} \delta^{cd} +
A(t|s,u) \delta_{a}^c \delta_b^{d}+A(u|s,t) \delta_{a}^d \delta_b^{c}\,,
\label{genericparam}
\eeq
where $s,t,u=-s-t$ are the usual Mandelstam invariants.
A 1-loop computation gives the amplitude up to four powers of momenta \cite{Weinberg:1978kz, Gasser:1983yg},
\beq
A(s|t,u)=\frac{s}{f^2_\pi}+\frac{1}{f^4_\pi}\left[\alpha s^2+ \beta (t^2+u^2)-\frac{s^2}{32 \pi^2 } \log{\frac{-s}{f^2_\pi}}-\frac{t-u}{96 \pi^2 }\left(t \log{\frac{-t}{f^2_\pi}}-u \log{\frac{-u}{f^2_\pi}}\right)\right]+\dots
\label{lowenansatz}
\eeq
The dimensionless parameters $\alpha$ and $\beta$ can be related to the Wilson coefficients $\ell_1$ and $\ell_2$.
However, this involves a choice of renormalization scheme and for this reason from now on we will always refer to the parameters $\alpha$ and $\beta$ which are defined by the equation above in terms of the physical scattering amplitude.
In appendix \ref{1loopAp}, we derive \eqref{lowenansatz} as the unitarity completion of the tree-level term $A(s|t,u)=\frac{s}{f^2_\pi}+\dots $. This makes transparent the fact that the coefficients of the logs in  \eqref{lowenansatz} are fixed in terms of $f_\pi$ and the polynomial part involves free parameters. We also do this exercise to next order in appendix \ref{2loopAp}. This gives the explicit form of the pion scattering amplitude up to six powers of momenta, in perfect agreement with the chiral limit of the 2-loop computation in \cite{Bijnens:1995yn,Bijnens:1997vq}, see appendix \ref{chiPTappendix}.

\begin{figure}[h]
  \centering
    \includegraphics[width=1 \textwidth]{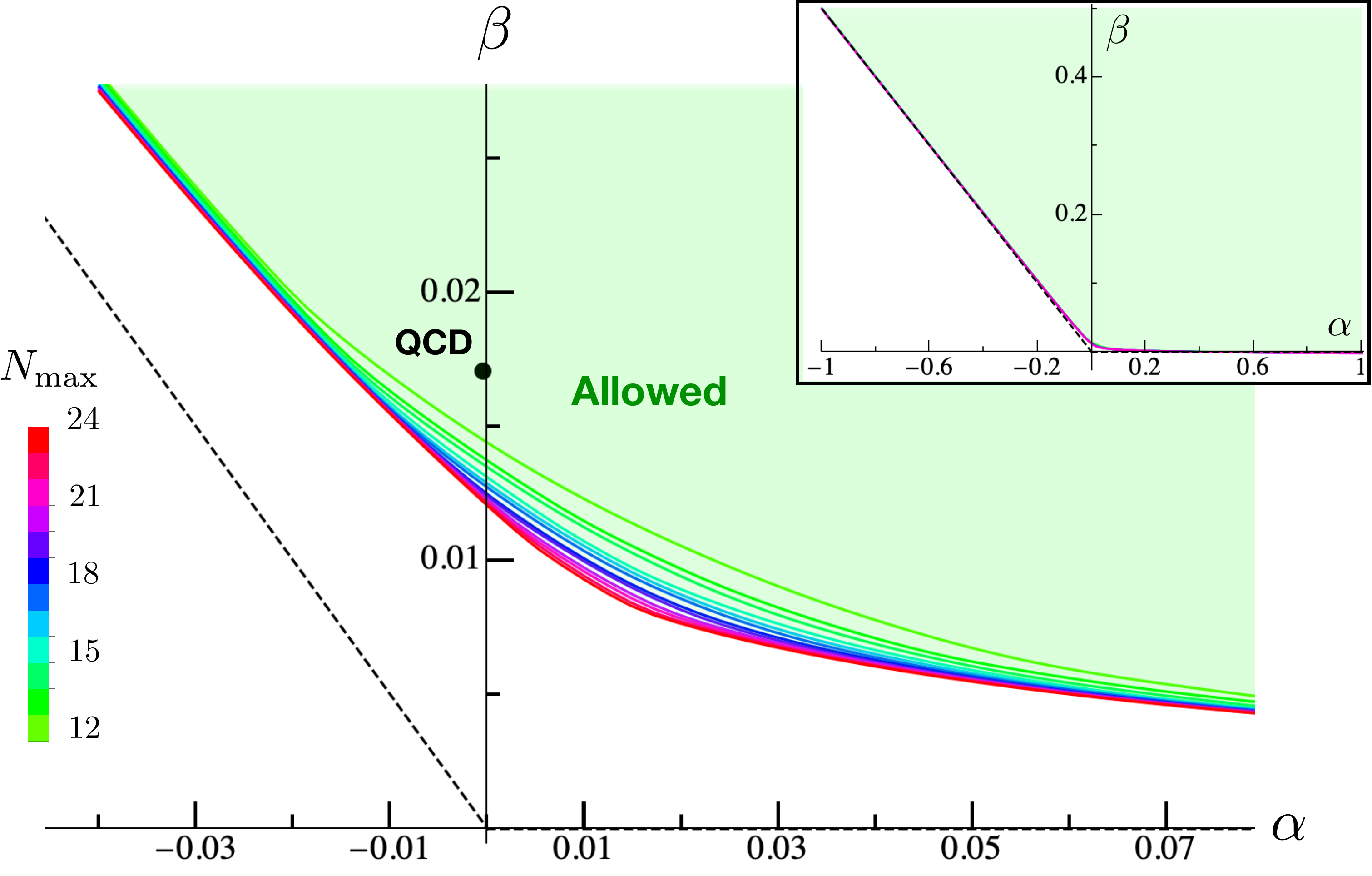}
  \caption{Allowed region in the $\{\alpha,\beta\}$ space for different values of $N_{\text{max}}$ ranging from 12 to 23.  For each $N_{\text{max}}$ we bound all the partial waves up to spin $L_{\text{max}}=90$ such that the spin cutoff dependence is negligible. In the inset it is shown a zoom out of the $\{\alpha,\beta\}$ space. The green region is allowed.
  The dashed lines denote the naive bounds \eqref{naivebounds} obtained from dispersion relations and unitarity, neglecting the effect of logarithmic branch cuts in the scattering amplitude. 
  Our numerical bound has a non-trivial shape in the natural range $\alpha \sim \beta \sim \frac{1}{(4\pi)^2}$ expected from naive dimensional analysis.
  Using values from \cite{Bijnens:1995yn,Bijnens:1997vq}, it seems that QCD lies inside but close  to the boundary of the allowed region. 
 } \label{fig1}
\end{figure}

In section \ref{sec:numerics}, we explain how the numerical S-matrix bootstrap can be used to derive universal bounds on the coefficients $\alpha$ and $\beta$. Our method relies only on the principles of Lorentz invariance,  crossing symmetry and unitarity and the assumption of Mandelstam analyticity of the scattering amplitude.
The allowed region in the $\{\alpha,\beta\}$ plane is depicted in figure \ref{fig1}.\footnote{Note that our results apply more broadly and describe any $O(3)$ symmetric theories with soft low energy behaviour in the $2\to 2$ amplitude. To distinguish between such generic theories and those arising from a particular symmetry breaking setup -- such as chiral symmetry breaking leading to the chiral Lagrangian -- we would need to consider higher point amplitudes or analyze the leading inelasticity of the $2\to 2$ process which only kicks in at very high loop order, at order $s^4$. Would be interesting to develop our analysis further to see how special are chiral symmetry breaking theories in the landscape of theories with a good soft behaviour.}
As explained in detail in section \ref{sec:numerics}, our method involves a parameter $N_\text{max}$ that controls the freedom of our ansatz for the amplitude. In figure \ref{fig1}, one can see that the allowed region is mostly stable when we increase $N_\text{max} \gtrsim 18$ except in a small region close to the origin where convergence is slower.
Remarkably, the empirical values of $\alpha$ and $\beta$ in real QCD are not too far from the boundary of the allowed region (see appendix \ref{chiPTappendix} for details).

\begin{figure}[h]
  \centering
    \includegraphics[width=1 \textwidth]{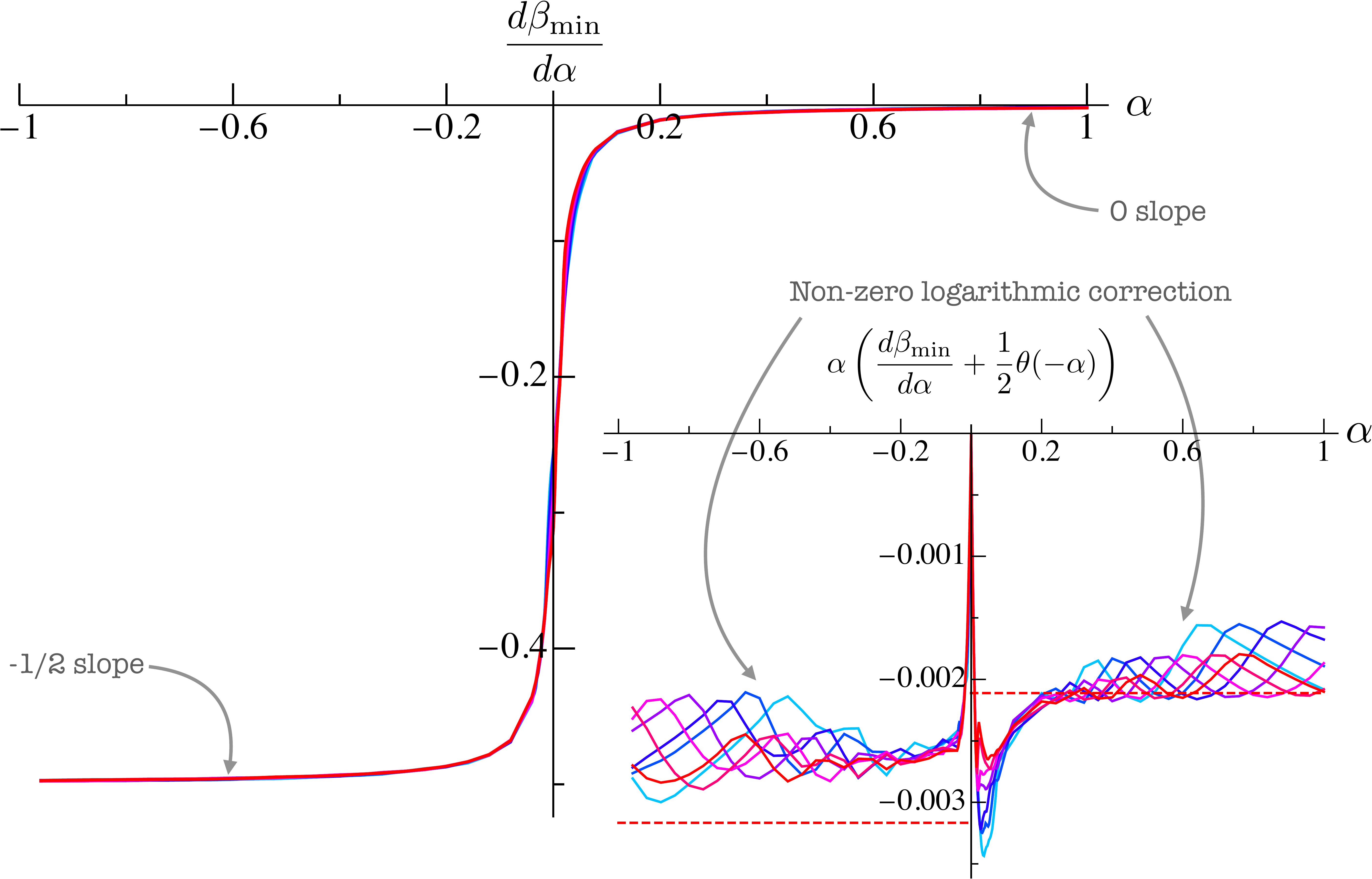}
  \caption{Discrete derivative of $\beta_\text{min}(\alpha)$ with respect to $\alpha$ for different values of  $N_\text{max}$ (same colour coding as in figure \ref{fig1}). We see that for large negative (positive) $\alpha$ we approach the expected $-1/2$ ($0$) slope from the naive bounds \eqref{naivebounds}.
 This leading linear behavior is already seen clearly in figure \ref{fig1}. Interestingly, there are important logarithmic corrections to this as illustrated in the inset. There, we plotted the combination $\alpha (d\beta_\text{min}/d\alpha + 1/2 \theta(-\alpha))$, involving the Heaviside $\theta$-function. This combination kills the leading linear asymptotics and extracts the coefficient of any log at large $|\alpha|$. Of course, numerical errors 
 are magnified when we multiply by $\alpha$ which is why we get the numerical oscillations in the inset.  
 Nevertheless, we see reasonable agreement with the analytic prediction \eqref{improvednaivebounds} shown as red dashed horizontal lines.  
 }
 \label{fig2}
\end{figure}

The inset of figure  \ref{fig1} suggests that the allowed region asymptotes to 
\beq
\la{naivebounds}
\beta >0 \quad \wedge \quad \alpha +2\beta > 0 \,,
\eeq
for large values of $\alpha$.
In fact, these naive bounds follow from the dispersive arguments of \cite{Pham:1985cr, Adams:2006sv} applied to  forward scattering amplitudes if we neglect the fact that the logarithmic branch cuts extend to $s=0$. We review this argument in appendix \ref{ap:arcs} and upgrade it to take into account the full non-perturbative analytic structure of the amplitude.
In figure \ref{fig2}, we make a more careful comparison between our numerical results and the naive formula \eqref{naivebounds}.
Firstly, we observe that the derivative $\frac{d\beta_\text{min}}{d\alpha}$ of the numerical bound asymptotes to 0 for $\alpha \to +\infty$ and to $-\frac{1}{2}$ for $\alpha \to -\infty$, in agreement with 
\eqref{naivebounds}.
Secondly, in the inset of figure  \ref{fig2}, we observe non-zero $\frac{1}{\alpha}$ corrections in $\frac{d\beta_\text{min}}{d\alpha}$ at large $|\alpha|$, which imply logarithmic corrections to the naive bounds \eqref{naivebounds}.
The dispersive analysis in appendix  \ref{ap:arcs} suggests the following asymptotic behavior of the allowed region
\begin{align}
\la{improvednaivebounds}
\beta &\ge -\frac{1}{48\pi^2} \log |\alpha| + \mathcal{O}(\alpha^0) \,,\qquad \qquad \alpha \to +\infty\,,\\
\alpha +2\beta &\ge -\frac{1}{16\pi^2} \log |\alpha|  + \mathcal{O}(\alpha^0) \,,\qquad \qquad \alpha \to -\infty\,.
\nonumber
\end{align}
This prediction is shown in red dashed lines in the inset of figure  \ref{fig2}. It works rather well for positive $\alpha$ and not so well for negative $\alpha$. It is unclear if this is simply due to numerical uncertainties or if the scenario proposed in appendix  \ref{ap:arcs} is not realized for large negative $\alpha$.

This short article provides a proof of concept that the numerical S-matrix bootstrap can be used to derive universal bounds on EFTs in more than two spacetime dimensions. 
A previous example in two dimensions is the flux tube S-matrix bootstrap \cite{FluxTube}.
In principle, the same methods can be applied to other EFTs describing other massless particles like photons or gravitons.\footnote{For gravitons, one needs to consider spacetime dimension $\ge 5$ to have well defined scattering amplitudes (without IR divergences).}
In particular, one can derive bounds to the leading higher curvature corrections to supergravity in 10 and 11 dimensions and compare them to the predictions of string theory \cite{SUGRA}.

\section{Numerics and some beautiful phase shifts}
\la{sec:numerics}
A big part of the setup, amplitude ansatz and numerics for massless pions follows almost verbatin the massive pion amplitude case studied in \cite{Andrea} which in turn was strongly based on the general
$\rho$ series parametrization of higher dimensional scattering amplitudes proposed in \cite{Paper3}. We assume familiarity with those ideas and will now highlight what is special to the case at hand. 

Following \cite{Paper3} we will think of the amplitude $A(s|t,u)$ as if it were a function of three independent variables $s,t$ and $u$. This is of course not true since $s+t+u=4m_\pi^2=0$ so we are \textit{extending} the two dimensional physical space manifold into an off-shell three dimensional bigger manifold. Nice mathematical properties of the larger manifold and of the sub-manifold guarantee that this analytic extension exists without the need to introduce any further singularities in the bigger space \cite{Paper3}.  
 
Then, we will   map the full complex plane for each of these three variables -- minus their two particle cuts at the positive real axis $\mathbb{R}^+$ -- to a unit disk. In the massive pion case we centered the disk at $s_*=t_*=u_*=4m_\pi^3/3$ which was a particularly nice point as it obeys the physical constraint $s+t+u=4m_\pi^2$. This is not longer a good expansion point now as it would collide with the two particle cut in the massless case. So instead we will map an arbitrary negative $s^*$ point to the centre of the unit disk. We picked $s_*=t_*=u_*= -16 f_\pi^2 $ and fixed units such that $f_\pi = \frac{1}{4}$.\footnote{The numerics converged better for this choice  than for the choice $s_*=t_*=u_*= - f_\pi^2 $. One possible explanation is that the mass  of the $\rho$-resonance $M_\rho \sim 8 f_\pi$ is a better estimate than $f_\pi$ for the characteristic energy scale in pion scattering amplitudes. } 
Then, one can write the analytic map as follows
\beq
\rho(s) \equiv \frac{1-\sqrt{-s}}{1+\sqrt{-s}}\,.
\eeq
Note  that $s_*+t_*+u_* \neq 0$ and therefore we are really now doing something conceptually different to the massive case: we are expanding around an off-shell point in the larger three dimensional $s,t,u$ enlarged space! The mathematical theorems quoted in \cite{Paper3} (about the vanishing of higher cohomologies of coherent analytic sheaves)
are thus more important than ever here. 

The $\rho$ variables map the low energy expansion point $s=0$ to $\rho=1$ and the high energy point $s=\infty$ to $\rho=-1$. This means it is particularly trivial to "unitarize" expressions by promoting them to rho variables. For example, $(\rho(s)-1)^2$ will behave as $s$ at small energies and will approach a constant at large $s$. Of course, we can build polynomials in $\rho$ which will approach any desired low energy behaviour without exploding at $s\to \infty$ since this high energy point simply corresponds to replacing $\rho\to -1$ in these polynomials.\footnote{We can even build decaying expressions easily if needed by multiplying these polynomials by appropriate powers of $\rho+1$.} 
Along these lines, we define
\bea
&&\!\!\!\!\! \texttt{low energy} \equiv-\frac{\chi(s)}{f_\pi^2}+\frac{1}{f_\pi^4}\left[\left(\alpha-3+\frac{\log{f_\pi^2}}{48 \pi^2}\right)\chi(s)^2+\left(\beta+\frac{\log{f_\pi^2}}{48\pi^2}\right)(\chi(t)^2+\chi(u)^2)\right.\nonumber\\
&&\qquad\qquad\qquad \left.-\frac{3\chi(s)^2 \log \chi(s)+(\chi(t)-\chi(u))(\chi(t)\log \chi(t)-\chi(u)\log\chi(u))}{96\pi^2} \right] \,,  
\la{lowEansatz}
\eea
with
\beq
\chi(s)\equiv \frac{1}{4}(\rho(s)-1)^2+\frac{1}{4}(\rho(s)-1)^3=-s-3 s^2+\mathcal{O}(s^{5/2}) \,,
\eeq
which matches precisely (\ref{lowenansatz}) at small $s,t,u$ but which is finite at high energy.  

Then our ansatz is simply 
\beq
A(s|t,u)=\texttt{low energy}+ { \sum' c_{ab}\rho(s)^a(\rho(t)^b+\rho(u)^b) + \sum' d_{ab}(\rho(t)^a\rho(u)^b+\rho(t)^b\rho(u)^a)} \la{ansatz}
\eeq
where the low energy behavior is automatically input. To precise that, and clarify a few other important points, we should explain what the primes in these sums stand for.

We have (for $s\to 0^-$ say), 
\beq
\rho(s) = 1-2\sqrt{-s}+ 2(-s) -2(-s)^{3/2} +2(-s)^2 + \dots
\eeq
which approaches $1$ as explained above. Note however that in the expansion around $1$ there are annoying half integer powers. This is of course unavoidable as   we are mapping the cut plane to the unit disk and we are thus opening up a square root cut. However, massless pion amplitudes do not have such square roots in their low energy expansion, see e.g. (\ref{lowenansatz}) or the two loop counterpart in appendix \ref{2loopAp}. Instead, we have logs as explicitly added by hand in (\ref{lowEansatz}). So how to get rid of these square roots? Well, at low energy this is easy: We write $t=(1+x)s/2$, $u=(1-x)s/2$ and expand the sums in (\ref{ansatz}) at small $s$. Then we fix appropriate linear combinations of $c_{ab}$ and $d_{ab}$ in (\ref{ansatz}) such that \textit{all} powers of $s$ in these sums (integers of half integers) cancel up to order $s^2$. This eliminates many constants but still leaves a huge amount of free parameters, see table \ref{tableCounting}. Since we killed all these terms in the low energy expansion we not only got rid of these annoying square roots but we also guaranteed that the full ansatz (\ref{ansatz}) perfectly captures the low energy behavior in (\ref{lowenansatz}). The primes in (\ref{ansatz}) thus stand simply for the sums truncated up to some large cutoff so that $a+b\le N_\text{max}$  and with the constants $c_{ab}$ and $d_{ab}$ simplified according to all the linear constraints generated up to~$O(s^2)$.\footnote{
Of course, we could go on and kill all $s^{5/2}$ half integer powers and even $s^{7/2}$ and so on. 
This might not be optimal.
After all, we expect more logs showing up at larger powers of~$s$ and those are not built in the ansatz either. 
An ansatz with more free parameters can better mimic log behaviors at higher powers of $s$.
This seems to be indeed what we observe numerically. 
Once we impose the absence of $s^{5/2}$ terms in the ansatz, the numerics seem to converge to the very same bounds as the ones described in the main text but convergence is slower. 
}

 \begin{table}[t]
  \begin{center}
    \begin{tabular}{c|c|c|c|c|c|c} 
$N_\text{max}$ & $s^0$ &  $s^{1/2}$ &  $s^1$ &  $s^{3/2}$ & {\color{blue} $s^{2}$} & $s^{5/2}$ \\
\hline
10&       89 &     88 &    85 & 82 & {\color{blue}79} & 74\\
12&     125 &    124 &    121 & 118 & {\color{blue}115} & 110\\
14&     167 &    166 &    163 & 160 & {\color{blue}157} & 152\\
16&     215 &    214 &    211 & 208 & {\color{blue}205} & 200\\
18&     269 &    268 &    265 & 262 & {\color{blue}259} & 254\\
20&     329 &    328 &    325 & 322 & {\color{blue}319} & 314\\
22&     395 &    394 &    391 & 388 & {\color{blue}385} & 380\\
24&     467 &   466 &    463 & 460 & {\color{blue}457} & 452
    \end{tabular}
        \caption{Number of parameters in our ansatz as we fix the proper low energy behavior to higher and higher orders in small $s$. 
        The column $s^a$ corresponds to setting to zero all terms in \eqref{ansatz} that greater or equal than $s^a$ in the small $s$ expansion.
        The plots in this paper are obtained for the next to the last column, in blue.}
    \label{tableCounting}
  \end{center}
\end{table}

  \begin{figure}[t]
  \centering
    \includegraphics[width=1 \textwidth]{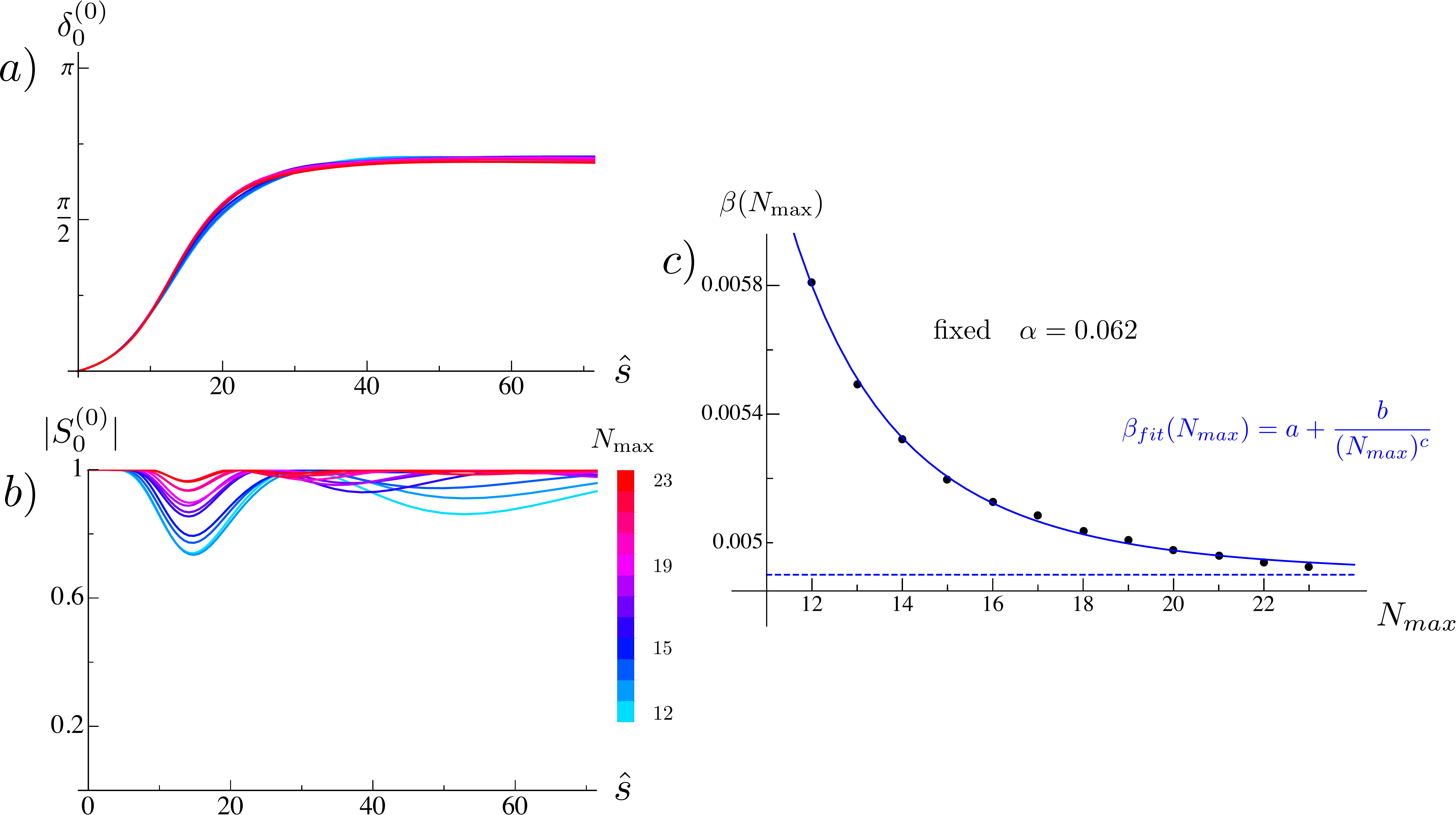}
\caption{(a) Example of a phase shift $\delta_{0}^{0} =\text{Re}( \frac{1}{2i} \log S_0^{(0)})$ at a given point at the boundary (corresponding to $\alpha =0.062$, spin $\ell=0$ and isospin $I=0$) as a function of $\hat s=s/f_\pi^2$ as we increase $N_\text{max}$. The phase shift seems to have converged to a nice resonance like shape. (b) The absolute value $|S_{0}^{(0)}|$ approaches $1$ more and more as we increase $N_\text{max}$. It clearly seems as if unitarity \textit{wants} to be saturated in the infinite $N$ limit. We also observe that this unitarity saturation is often achieved at the price of some more erratic behaviour at high energy which we do not control well. Given the interesting interplay between unitarity and the Aks theorem \cite{aks1,aks2} as recently emphasized in \cite{Correia:2020xtr}, it would be very interesting to study this unitarity (non)-saturation in much more detail. (c) To reach the optimal infinite $N$ bound we can try to fit our optimal target (in this case we are minimizing $\beta$ at fixed $\alpha$) as we increase $N_\text{max}$ and try to extrapolate. 
} \label{fig4}
\end{figure}

 \begin{figure}[t!]
  \centering
    \includegraphics[width=0.78 \textwidth]{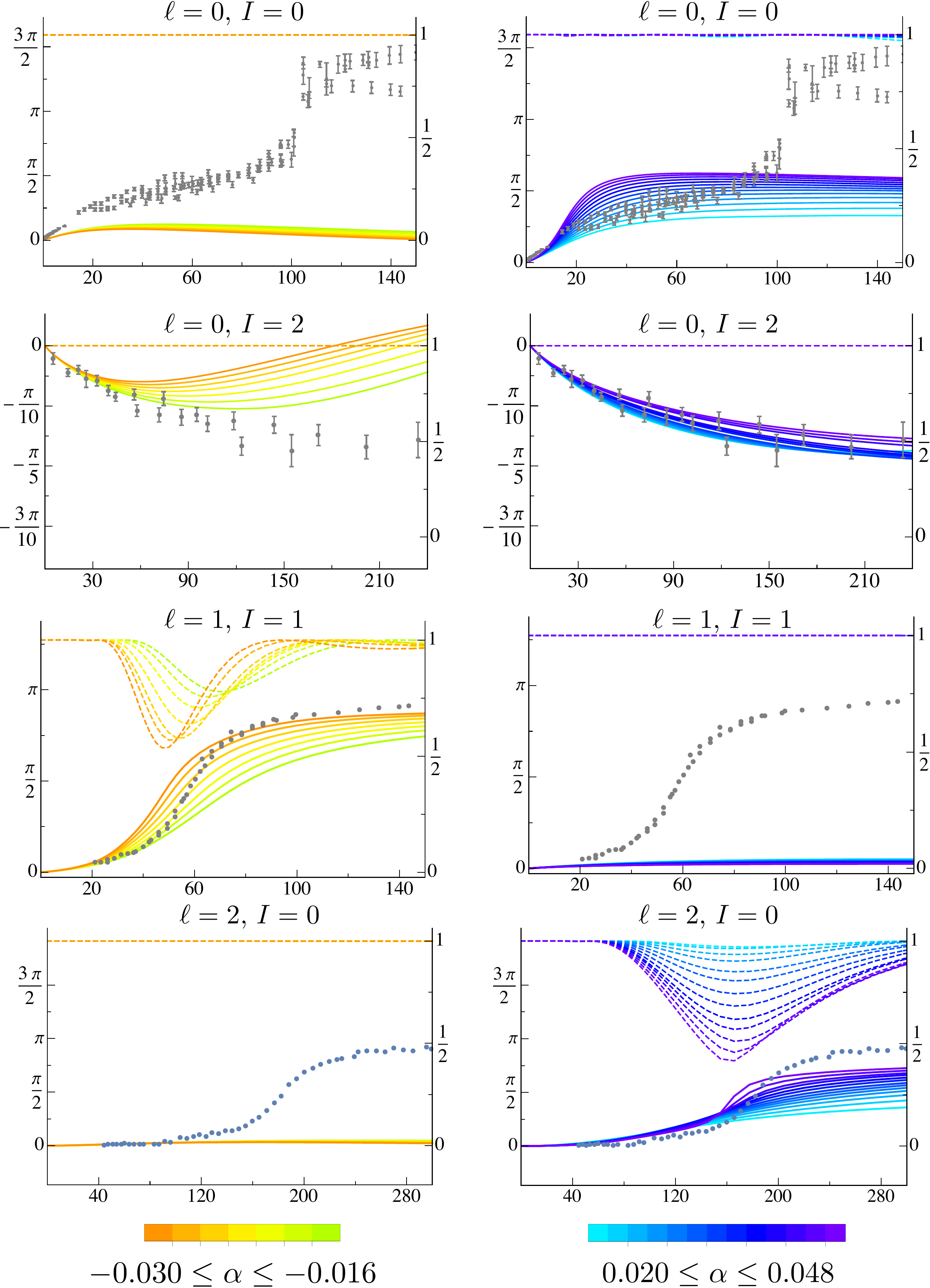}
\caption {Phase shifts $\delta_{l}^{I} =\text{Re}( \frac{1}{2i} \log S_l^{(I)})$ (solid lines; left y-axis)   and absolute values of the corresponding amplitudes (dashed lines; right y-axis) for the lowest spins and isospins along the boundary as function of $\hat s=s/f^2_\pi$, with $N_\text{max}=23$. 
The gray dots and error bars are experimental phase shifts for real world massive pions \cite{Protopopescu:1973sh,Losty:1973et,Grayer:1974cr,Estabrooks:1974vu,Hoogland:1977kt,Batley:2010zza,GarciaMartin:2011cn}. 
\protect\footnotemark~The comparison is done by plotting the experimental data also 
as a function of the square of the center of mass momentum in units of $f_\pi$, i.e.  
$(s-4m_\pi^2)/f_\pi^2$. Here we use $m_\pi\simeq 140\text{Mev}$ and $f_\pi\simeq 93 \text{MeV}$.
}
  \label{panels}
\end{figure}
\footnotetext{We thank Jos\'e Ram\'on Pel\'aez for sharing with us the set of available experimental data on $\pi-\pi$ scattering.}

This basically concludes the highlights particular to our massless particles setup. The next steps can be taken from the massive analysis \cite{Paper3} and \cite{Andrea} by setting $m=0$ there. For example, formula (2) in \cite{Andrea} for the pion amplitude decomposition into partial waves simply becomes (note that $4=4m_\pi^2$ there)
\begin{eqnarray}
\( \begin{array}{c} 
3A(s|t,u)+A(t|s,u)+A(u|s,t)\\ 
A(t|s,u)-A(u|s,t) \\
A(t|s,u)+A(u|s,t) \end{array} \) = 16\pi i  \sum_{{l}} (2l+1) P_l\Big(\frac{u-t}{u+t}\Big) \( \begin{array}{c} 
1- {{S_l^{(0)}(s)}}\\ 
1- {{S_l^{(1)}(s)}}\\
1- {{S_l^{(2)}(s)}}\end{array} \) \label{decomposition}
\end{eqnarray}
and unitarity is simply 
\begin{eqnarray}
|S_l^{(I)}| \le 1 \,, \qquad s>0 \,.  \la{unit}
 \end{eqnarray}
So at this point the rules of the game are pretty much as in \cite{Paper3} and \cite{Andrea}: We pick the ansatz~(\ref{ansatz}) with a cut-off~$N_\text{max}$ in the sums, construct the partial amplitudes $S_l^{(I)}$  through~(\ref{decomposition}) from spin $0$ up to some large spin $L_\text{max}$ and pick an $s$ grid with $M_\text{grid}$ points where to impose~(\ref{unit}) for each   spin $l$ and isospin $I$. Then, with these many parameters and constraints we explore the allowed $(\alpha,\beta)$ space as introduced in (\ref{lowenansatz}). (For example, we can pick some $\alpha$ and minimize $\beta$ as illustrated for $\alpha \simeq 0.06$ in figure \ref{fig4}). 
If  the  parameters  $N_\text{max}$, $L_\text{max}$ and $M_\text{grid}$ are large enough things should converge.

In this problem we also found very important to impose the large spin constraints obtained by estimating analytically the large spin partial waves following 
\cite{Paper3}. In the previous massive explorations this large spin analytic constraints would be irrelevant in practice as we would reach the asymptotic large spin regime well before the cut-off $L_\text{max}$. This is expected; massive particles mediate short range interactions so large spin, corresponding to large impact parameter, quickly becomes exponentially subleading. This is not so for massless particles where this suppression becomes power like. So here the large spin constraints are crucial to ensure proper convergence. (For more details see appendix D.4 of \cite{Paper3}).

Even with those large spin conditions and even for quite large $N_\text{max}$, these numerics are very hard and things converge well but not splendidly as illustrated in the introduction, see figure \ref{fig1}. 
(For the highest $N_\text{max}=23$ we have used, there are 420 free parameters, $18\times 10^3$ quadratic unitarity constraints, 200 linear higher spin constraints.\footnote{For the grid we use a Chebyschev grid in $\rho$ (unit disc boundary) with 200 points for spins from $\ell=0$ to $\ell=49$ and 50 points for spins above $\ell=50$. There are also the 200 higher spin constraints. This leads to the number of   quadratic unitarity constraints quoted in the text.} We have used SDPB Elemental \cite{Elemental} on 40 cores and it took 6h 40m per optimization problem, that is per point at the boundary of figure \ref{fig1}.) 

\begin{figure}[t!]
  \centering
    \includegraphics[width=1
     \textwidth]{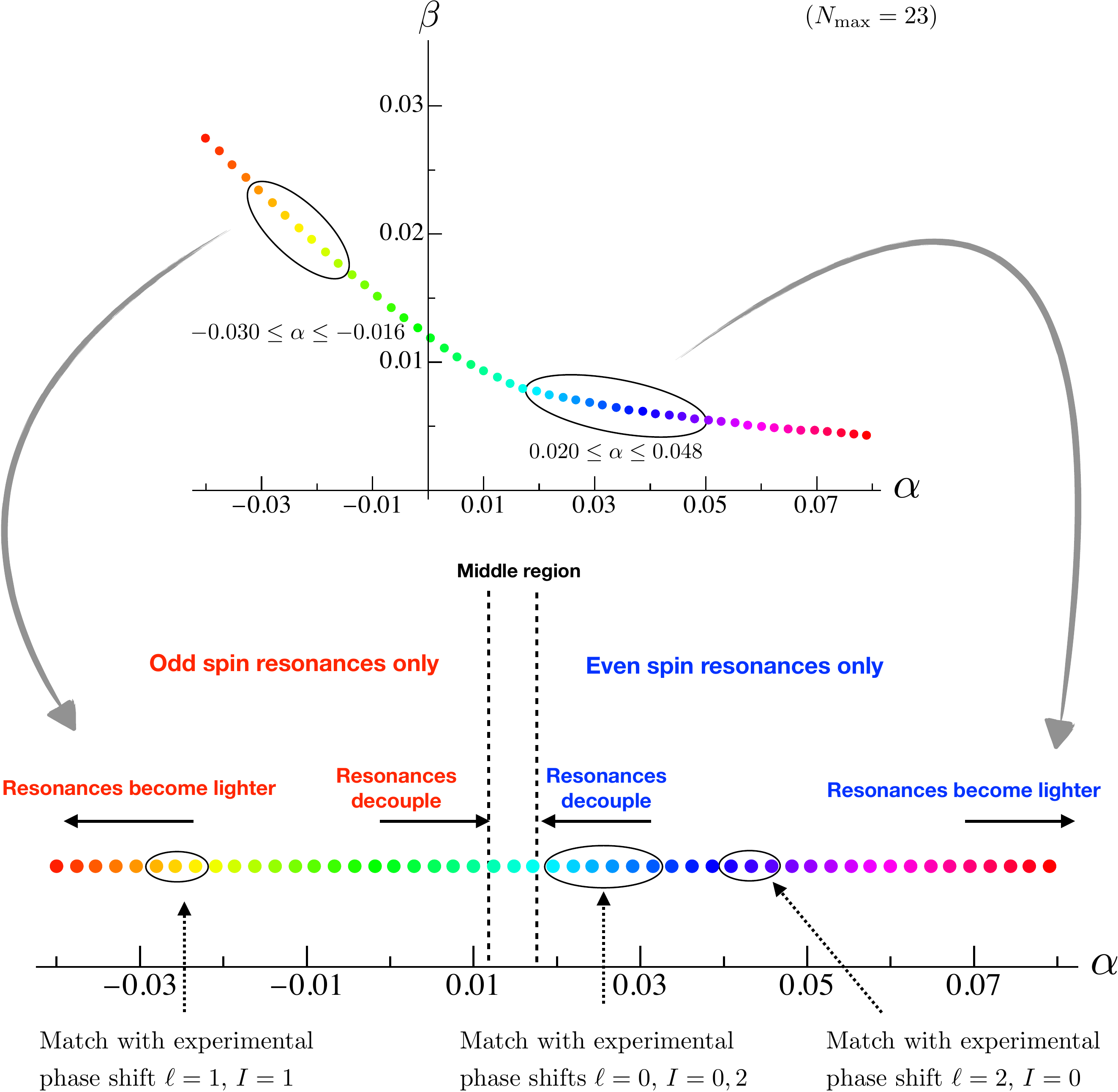}
  \caption{There is a rich pattern of resonances in the optimal S-matrices as we move along the allowed S-matrix space. On the left, we have odd spin resonances turned on and no even spin resonances. On the right we have the opposite. The real world has both types of resonances so should somehow not be at the boundary. Would be very interesting to force the presence of the $\rho$ resonance as in \cite{Andrea} and repeat this analysis as done in that paper for massive pions. 
  }\label{fig5}
\end{figure}

It would be extremely useful to find a dual formulation of the S-matrix optimization problem considered here. This would be specially important in the slow convergence region of figure \ref{fig1} because it would give rigorous bounds even with finite truncations, i.e. it would approach the bounds \textit{from the opposite side}.
We are optimistic that such methods will be soon available building upon the recent developments
in the dual S-matrix bootstrap program \cite{Monolith, dualPaper, MartinTalkBootstrap}.

In the meantime we can take advantage of a major positive point of the original or primal formulation: We are explicitly constructing $S$-matrices compatible with unitarity and crossing symmetry so we can literally look at the optimal phase shifts as we move along the boundary of the allowed parameter space! This is a particularly exciting exercise in the case of pions where we have real world experimental data for their phase shifts \cite{Protopopescu:1973sh,Losty:1973et,Grayer:1974cr,Estabrooks:1974vu,Hoogland:1977kt,Batley:2010zza,GarciaMartin:2011cn}.\footnote{Of course, the attentive reader might point out that such comparison with the real world would perhaps make even more sense in the massive setup explored in \cite{Andrea, Bose:2020shm}. That is true. It would be very interesting to better explore the massless limit interpolation starting from a massive setup. In particular, starting with masses could be helpful in controlling the issue of log versus square root singularities discussed above.} The outcome of this exploration is depicted in figure \ref{panels} where we highlight two interesting regions along the boundary where the numerics do resemble nicely experimental data. 
On the right region, in blue, we see that the even spins/isospins actually seem to be quite similar to the experimental ones while spin $1$ partial wave is unfortunately quite off. In particular, the experimental data have a clear resonance around $s=30$ corresponding to the $\rho$ particle which is clearly missing in the blue amplitudes. On the left orange region it seems to be the opposite: we seem to have nice structure in the odd spin phase shifts which do resemble closely the experimental phase shifts but the even spin phase shifts are now quite off. Figure \ref{fig5} contains a graphical summary. In sum, without imposing further constraints, the real world does not seem to be close to the boundary. Of course, it did not need to be. It would be interesting to add further constraints to our numerics -- fixing the position of the $\rho$ particle for example -- and redoing all the bounds. Perhaps with this extra physical constraint the S-matrices at the boundary might approach those observed in experiment? Will they have rich even spin structures arising? We leave those numerical explorations to the future.


 \section*{Acknowledgements}
We would like to thank Gilberto Colangelo, Joan Elias-Mir\'o and Francesco Riva for interesting 
discussions and comments on the draft. We thank Alessandro Pilloni and Riccardo Rattazzi for illuminating discussions. 
We thank Jose Pelaez for interesting discussions and for sharing valuable pion data with us. 
Research at the Perimeter Institute is supported in part by the Government of Canada 
through NSERC and by the Province of Ontario through MRI. This work was additionally 
supported by a grant from the Simons Foundation (Simons Collaboration on the Nonperturbative Bootstrap: JP:\#488649 and PV: \#488661) and ICTP-SAIFR FAPESP grant 2016/01343-7 and FAPESP grant 2017/03303-1. The work of P. V.  was partially supported by the Sao Paulo Research Foundation (FAPESP) under Grant No. 2019/24277-8. 
JP is supported 
by the Swiss National Science Foundation through the project 200021-169132 and through the
National Centre of Competence in Research SwissMAP. 
AG was supported by The Israel Science Foundation (grant number 2289/18).



\appendix
\section{Perturbative Unitarity} \la{ap:PU}
The scattering of Goldstone particles is free of IR divergences. The corresponding amplitudes have a good soft behavior and vanish as the momenta at sent to zero. 

For a single goldstone particle, for example we would write a fully crossing symmetric amplitude low energy expansion $A(s,t,u) \simeq A (s+t+u) + B(s^2+t^2+u^2)+ C (s+t+u)^2+\dots$ but the $A$ and $C$ terms would drop due to $s+t+u=0$ and the first leading interaction starts at quadratic order and is governed by $B$. 

%

In our case we have an  $O(N)$ scattering amplitude in (\ref{genericparam})  is parametrized in terms of a single scalar function $A(s|t,u)$ symmetric in $t,u$ only.
The simplest and most general tree-level on-shell interaction can now start already at linear order as 
\beq
A^{\text{tree}}(s|t,u)=\frac{s}{\Lambda^2},
\label{treeseed}
\eeq
where $\Lambda=f_\pi$ in the case of massless pions.

In what follows we will solve this simple exercise: starting from the seed in eq. \eqref{treeseed}, we will make use of crossing and analyticity to write an ansatz for the imaginary part of the~$\mathcal{O}(s^2)$ or 1-Loop amplitude, and then impose elastic unitarity saturation to fix the coefficients of the ansatz.
We will be able to find the parametrization shown in eq. \eqref{lowenansatz} and repeating the exercise at the next order we will also determine the most general parametrization for the~$\mathcal{O}(s^3)$ or 2-Loops amplitude up to a few "theory dependent" coefficients. \footnote{The leading logarithms can be extracted at any loop order using  the non-linear recursion relations in \cite{Koschinski:2010mr}.}

At $\mathcal{O}(s^4)$ particle production kicks in: the $2\to 4$ process contributes to the imaginary parts of 3-Loops diagrams that would appear at this order.

\subsection{$O(N)$ scalar particles at 1-Loop}
 \la{1loopAp} 


In this section we will prove eq. \eqref{lowenansatz}.
In the massless case all the normal thresholds start at $s=t=u=0$, and we assume there are no other singularities in the complex plane of the Mandelstam variables.
The minimal ansatz for the 1-Loop amplitude must contain at least polynomials and logarithms at the order $\mathcal{O}(s^2)$.

Using crossing symmetry and setting $\Lambda=1$, we can write the expansion
\bea
&&A(s|t,u)=s+\alpha s^2+ \beta (t^2+u^2)+(a_1 s^2 +a_2 (t^2+u^2))\log{(-s)}+\nonumber\\
&&+(b_1 s^2 +b_2 (t^2+u^2))\log{tu}+b_3 (t^2-u^2)\log{\frac{t}{u}}+\mathcal{O}(s^3).
\label{pertansatz}
\eea
To impose unitarity we must project the ansatz \eqref{pertansatz} into irreps of the global symmetry group and partial waves.
Let's define the $s$-channel projections over the irreps of $O(N)$, respectively the singlet, antisymmetric and symmetric traceless reps
\begin{align}
\mathcal{T}^{(0)}(s,t,u)&=N A(s|t,u)+A(t|s,u)+A(u|s,t)\\
\mathcal{T}^{(1)}(s,t,u)&=A(t|s,u)-A(u|s,t)\\
\mathcal{T}^{(2)}(s,t,u)&=A(t|s,u)+A(u|s,t),
\end{align}
and the projections in partial waves
\beq
t_\ell^{(I)}(s)=\frac{1}{32\pi}\int_{-1}^1 dx\,P_\ell (x)\mathcal{T}^{(I)}(s,t(s,x),u(s,x)),
\eeq
where $t=-s/2(1-x)$ and $u=-s/2(1+x)$.

Elastic unitarity of the S-matrix translates into $|S_\ell^{(I)}(s)|^2=1 $ or using $S_\ell(s)=1+ i t_\ell(s)$ 
we obtain a condition in terms of the partial amplitudes
\beq
2\IM {t_\ell^{(I)}}=|t_\ell^{(I)}|^2
\label{unitarity}
\eeq
At leading order in the momentum expansion the equation above relates the imaginary part of the one loop terms to the square of the tree-level amplitude:
\beq
2 \IM t^{1-\text{Loop}}=|t^{\text{Tree Level}}|^2\,,
\eeq
for each spin and isospin. 

At the leading order
\bea
&&\mathcal{T}^{(0)}(s,t,u)=(N-1)s+\mathcal{O}(s^2)\nonumber\\
&&\mathcal{T}^{(1)}(s,t,u)=t-u+\mathcal{O}(s^2)\nonumber\\
&&\mathcal{T}^{(2)}(s,t,u)=-s+\mathcal{O}(s^2),
\label{irreps}
\eea
where we explicit show their tree-level expressions.
Only the terms of order $\mathcal{O}(s^2)$ of \eqref{pertansatz} contribute to the imaginary part of the amplitude.
Using the replacement rule $\log(-s)=\log(s)-i\pi$ we get 
\bea
&&\IM \mathcal{T}^{(0)}(s,t,u)=-\pi (N(s^2 a_1+(t^2{+}u^2)a_2)+(t^2{+}u^2)(b_1+b_2-b_3)+2s^2(b_2{+}b_3))\nonumber\\
&&\IM \mathcal{T}^{(1)}(s,t,u)=\pi \,s (t-u)(b_1-b_2+b_3)\nonumber\\
&&\IM \mathcal{T}^{(2)}(s,t,u)=-\pi ((t^2+u^2)(b_1+b_2-b_3)+2s^2(b_2+b_3))).
\eea
Since the imaginary part is just a polynomial in the Mandelstam's of degree 2 we will only have non vanishing partial amplitudes up to spin 2.

Finally, we have a set of 5 equations
\bea
 &&\ell=0,\,I=0\,\implies -\frac{s^2}{48}(3 N a_1+2N a_2+2 b_1+8 b_2+4 b_3)=\frac{1}{2}\left(\frac{(N-1)s}{16\pi}\right)^2\nonumber\\
 &&\ell=2,\,I=0\,\implies -\frac{s^2}{240}(N a_2+b_1+b_2-b_3)=0\nonumber\\
 &&\ell=1,\,I=1\,\implies \frac{s^2}{48}(b_1-b_2+b_3)=\frac{1}{2}\left( \frac{s}{48\pi } \right)^2\nonumber\\
 &&\ell=0,\,I=2\,\implies -\frac{s^2}{24}(b_1+ b_2-b_3)=\frac{1}{2}\left(-\frac{s}{16\pi }\right)^2\nonumber\\
 &&\ell=2,\,I=2\,\implies -\frac{s^2}{240}(b_1+b_2-b_3)=0,
\eea
whose solution yields 
\beq
a_1=-\frac{N-2}{32 \pi^2},\quad a_2=0,\quad b_1=-b_3=\frac{1}{192 \pi^2},\quad b_2=-\frac{1}{96\pi^2},
\eeq
and
\bea
A(s|t,u)&&=s+\alpha s^2+ \beta (t^2+u^2)\nonumber\\
&&-\frac{N-2}{32 \pi^2} s^2 \log{(-s)}-\frac{1}{96 \pi^2}(t-u)(t\log{(-t)}-u\log{(-u)})+\mathcal{O}(s^3).
\eea
For $N=3$ we recover eq.~\eqref{lowenansatz}. As a check, in appendix~\ref{chiPTappendix} we recover directly eq.~\eqref{lowenansatz} taking the massless limit 
of the one-loop $\pi \pi$ scattering computation in~\cite{Bijnens:1995yn,Bijnens:1997vq}.

\subsection{$O(N)$ scalar particles at 2-Loops} \la{2loopAp}

Perturbative unitarity at the next to leading order 
\beq
\IM t^{\text{2-Loops}}=\RE t^{\text{1-Loop}} \times t^{\text{Tree}}.
\label{unitarity2loops}
\eeq
suggest that we have to include into the ansatz functions whose imaginary part is a logarithm.

We write the most generic crossing symmetric ansatz at $\mathcal{O}(s^3)$ include $\log^2$ as 
\bea
&&A(s|t,u)^{\text{2-loops}}=\nonumber\\
&&\gamma s^3+\delta (t^3+u^3)+\log{(-s)}(c_1 s^3+c_2 (t^3+u^3))+\log(t u)(c_3 s^3+c_4 (t^3+u^3))\nonumber\\
&&+\log{\frac{t}{u}}(c_5 s^2(t-u)+c_6 (t^3-u^3))+\log^2{(-s)}(d_1 s^3+d_2 (t^3+u^3))\nonumber\\
&&+(\log^2{(-t)}+\log^2{(-u)})(d_3 s^3+d_4 (t^3+u^3))+\log{(-t)}\log{(-u)}(d_5 s^3+d_6 (t^3+u^3))\nonumber\\
&&+(\log^2{(-t)}-\log^2{(-u)})(d_7 s^2(t-u)+d_8 (t^3-u^3))+\log{(-s)}\log{t u}(d_9 s^3+d_{10} (t^3+u^3))\nonumber\\
&&+\log{(-s)}\log{\frac{t}{u}}(d_{11} s^2(t-u)+d_{12} (t^3-u^3)).
\label{2loopansatz}
\eea

Following the 1-Loop analysis, we project eq. \eqref{2loopansatz} onto irreps of $O(N)$ and partial waves and solve the unitarity equation \eqref{unitarity2loops} in each channel.
Since the imaginary part now contains logarithms, the spin projections do not truncate and in principle we have an infinite system of linear equations. 
However, the right-hand side of eq. \eqref{unitarity2loops} do truncate. Therefore, at higher spin we only get homogeneous equations which turn out to be proportional to each other, hence, we can solve for the coefficients $c_i$ and $d_i$.

To compute the projections we need to evaluate integrals of the form
\beq
\int_{-1}^1dx\, x^a \log^b{\left(\frac{s}{2}(1-x)\right)}\log^c{\left(\frac{s}{2}(1+x)\right)}.
\eeq
They can be analytically evaluated performing the simple change of variables $x\to 2x-1$
and using the identity
\beq
\int_0^1dx\,x^a \log^m{x}\log^n{(1-x)}=\frac{\p^m}{\p\mu^m}\frac{\p^n}{\p\nu^n}\text{Beta}(a+\mu+1,\nu+1)\biggr|_{\mu=\nu=0}.
\eeq

The final answer for the 2-Loops amplitude is then completely fixed by elastic unitarity saturation 
\bea
&&A(s|t,u)^{\text{2-loops}}=\gamma s^3+\delta (t^3+u^3)+\frac{N(9N-20)+19}{9216\, \pi^4}s^3\log^2{(-s)}\la{eq30}\\
&&-\frac{1}{18432 \pi^4}((3N{+}11)(t^3\log^2{(-t)}{+}u^3\log^2({-u})) -6(N{-}3)tu(t\log^2{(-t)}{+}u\log^2{(-u)}))\nonumber\\
&&-\left((3N-1)\frac{\alpha}{48\pi^2}+(N+3)\frac{\beta}{24\pi^2}+\frac{11N-10}{27648\,\pi^4}\right)s^3\log{(-s)}\nonumber\\
&&+\frac{\alpha}{96\pi^2}(t^2(t{-}2u)\log{(-t)}+u^2(u{-}2t)\log{(-u)})\nonumber\\
&&+\frac{\beta}{96\pi^2}(t^2(2u{+}9t)\log{(-t)}+u^2(2t{+}9u)\log{(-u)})\nonumber\\
&&+\frac{1}{110592\pi^4}((21N{-}17)(t^3\log{(-t)}+u^3\log{(-u)})-2(3N{-}5)tu(t\log{(-t)}+u\log{(-u)})).\nonumber
\eea
At this order there are 2 new ``theory dependent" coefficients $\{\gamma, \delta\}$ parametrizing the on-shell polynomial deformations of order $\mathcal{O}(s^3)$. We could also explore their allowed space but the numerics would be quite a bit harder.

%
%
%
\section{Chiral limit of the massive result}
\la{chiPTappendix}

The one-loop analysis of the chiral perturbation theory has been carried out in all details long ago~\cite{Gasser:1983ky}.
In the modern literature the low energy constants are still defined in the same way as done in that pioneering work.
To help the phenomenology-oriented reader we write explicitly the match between our convention and the usual one.

We take the expression for the one-loop amplitude in the massive case from~\cite{Bijnens:1995yn,Bijnens:1997vq}, where the expansion has been worked out up to two-loops:
\beq
A(s|t,u)^{\text{1-loop}}=\frac{1}{f_\pi^4}(b_1 m^4+b_2 m^2+b_3 s^2+b_4 (t-u)^2)+\frac{1}{f_\pi^4}(F^{(1)}(s)+G^{(1)}(s,t)+G^{(1)}(s,u)),
\label{1loopchiral}
\eeq
where
\bea
&&F^{(1)}(s)=\frac{1}{2}(s^2-m^4)J(s)\\
&&G^{(1)}(s,t)=\frac{1}{6}(2 t^2+s t-10 t m^2-4 s m^2+14 m^4)J(t),
\eea
and
\beq
J(s)=\frac{1}{8\pi^2}\left(1+\frac{1}{2}\sqrt{z(s)}\,\log{\frac{\sqrt{z(s)}-1}{\sqrt{z(s)}+1}}\right),\qquad z(s)=\frac{s-4m^2}{s}.
\eeq
The first term in \eqref{1loopchiral} is just polynomial in the Mandelstam variables and it contains the running couplings of the counterterms for the one-loop divergences. 
The second term contains the universal functions fixed by unitarity with the 2-particles normal branch points.
We now take the chiral limit of the two terms separately.
When we take $m\to 0$ only the $b_3$ and $b_4$ constants survive. 
Their relations to the chiral Lagrangian couplings are given by
\bea
&&b_3=2 \ell^r_1(\mu)+\frac{1}{2}\ell_2^r(\mu)-\frac{1}{32 \pi^2}\log{\frac{m^2}{\mu^2}}-\frac{1}{16 \pi^2}\frac{7}{12}+\mathcal{O}\left(\frac{m^2}{f_\pi^2}\right),\\
&&b_4=\frac{1}{2}\ell_2^r(\mu)-\frac{1}{96 \pi^2}\log{\frac{m^2}{\mu^2}}-\frac{1}{16 \pi^2}\frac{5}{36}+\mathcal{O}\left(\frac{m^2}{f_\pi^2}\right).
\eea
Collecting all the terms we obtain
\bea
\frac{1}{f_\pi^4}(b_3 s^2+b_4 (t-u)^2)&=&\frac{s^2}{f_\pi^4}\left(2 \ell^r_1(\mu)-\frac{1}{36\pi^2}\right)+\frac{t^2+u^2}{f_\pi^4}\left( \ell_2^r(\mu)-\frac{5}{288\pi^2} \right)\nonumber\\
&&-\frac{s^2+t^2+u^2}{48\pi^2 f_\pi^4}\log{\frac{m^2}{\mu^2}}.
\eea
The massless limit of the unitarity logarithms is easily done noticing that
\beq
J(s)=\frac{1}{8\pi^2}\left(1-\frac{1}{2}\log\frac{-s}{m^2}\right)+\mathcal{O}\left(\frac{m^2}{s}\right).
\eeq
Replacing the leading behavior of $J(s)$ in the second term of \eqref{1loopchiral} yields
\bea
\frac{1}{f_\pi^4}(F^{(1)}(s)+G^{(1)}(s,t)+G^{(1)}(s,u))&=&\frac{1}{24 \pi^2 f_\pi^4}(s^2+t^2+u^2)-\frac{s^2}{32 \pi^2 f_\pi^4}\log\frac{-s}{m^2}\nonumber\\
&&-\frac{t-u}{96 \pi^2 f_\pi^4}\left(t \log{\frac{-t}{m^2}-u\log{\frac{-u}{m^2}}}\right).
\eea
The two terms are both logarithmically infrared divergent in the chiral limit. 
However, if we rescale the unitarity logarithms by $\log(-x/m^2)\to \log(-x/\mu^2)+\log(\mu^2/m^2)$ and we sum the two terms, then the final result is finite in the chiral limit
and coincides with eq.~\eqref{lowenansatz} if we fix the scale at $\mu=f_\pi$. (Similarly, the chiral limit of the two loop result in \cite{Bijnens:1995yn,Bijnens:1997vq} precisely matches with~(\ref{eq30}))
The relation between our parametrization and the coefficients in the chiral Lagrangian is given by
\beq
\alpha=2 \ell_1^r(f_\pi)+\frac{1}{72 \pi^2},\quad \beta=\ell^r_2(f_\pi)+\frac{7}{288\pi^2}.
\eeq

The numerical values for the $\ell_i^r$ quoted in \cite{Bijnens:1995yn,Bijnens:1997vq} are
\beq
\ell_1^r(M_\rho)=-5.40 \times 10^{-3},\qquad \ell_2^r(M_\rho)=5.67 \times 10^{-3}.
\eeq

We can obtain their values at $\mu=f_\pi$ using the RG equations 
\beq
16 \pi^2 \mu \frac{d \ell_1^r}{d\mu}=-\frac{1}{3},\qquad 16 \pi^2 \mu \frac{d \ell_2^r}{d\mu}=-\frac{2}{3},
\eeq
getting
\bea
&&\ell_1^r(f_\pi)=\ell_1^r(M_\rho)-\frac{1}{3}\frac{1}{16 \pi^2}\log\frac{f_\pi}{M_\rho}=-9.4 \times 10^{-4}\quad \implies \quad \alpha=-4.7 \times 10^{-4}\label{alphaqcd}\\
&&\ell_2^r(f_\pi)=\ell_2^r(M_\rho)-\frac{2}{3}\frac{1}{16 \pi^2}\log\frac{f_\pi}{M_\rho}=1.46 \times 10^{-2} \quad \implies \quad \beta=1.70 \times 10^{-2}\label{betaqcd},
\eea
where we have used $f_\pi=93 \,\text{MeV}$ and $M_\rho=770 \,\text{MeV}$.
In figure \ref{fig1} we compare the results in \eqref{alphaqcd}, and \eqref{betaqcd} with the allowed region in the $\{\alpha, \beta\}$ space.

\section{Asymptotic bounds from dispersion relations }
\la{ap:arcs}

In this appendix, we derive a dispersive representation of the parameters $\alpha$ and $\beta$ involving the imaginary part of two independent forward scattering amplitudes, which we define in section \ref{forwardamps}. The main argument, presented in \ref{arcsderivation}, leads to a prediction for the asymptotic behavior of the  lower bound on $\beta$ for large values of $|\alpha|$.
Throughout this appendix, we set units $f_\pi=1$.

\subsection{Crossing symmetric forward scattering amplitudes}
\la{forwardamps}

There are two crossing symmetric forward scattering amplitudes that we can define in pion physics.
Firstly, we can consider neutral pion scattering $\pi^0 +\pi^0 \to \pi^0+\pi^0$. This is described by the amplitude 
\beq
A(s|t,u)+A(t|s,u)+A(u|s,t)\,,
\eeq
which becomes
\beq
M_{\pi_0\pi_0}(s) \equiv A(s|0,-s)+A(0|s,-s)+A(-s|s,0)\,,
\eeq
in the forward limit.
Secondly, we can consider the transmission amplitude for the process $\pi^a +\pi^b \to \pi^a+\pi^b$ with $a\neq b$. The corresponding amplitude is 
\beq
A(t|s,u)\,,
\eeq
which in the forward limit reduces to
\beq
M_t (s) \equiv A(0|s,-s)\,.
\eeq

Clearly, both forward amplitudes $M_{\pi_0\pi_0}(s)$ and $M_{t}(s)$ are invariant under the crossing symmetry transformation $s \to u=-s$. This relates the values of the amplitude above and below the cut along the real axis in the $s$-complex plane. However, when combined  with real analyticity $M(s^*)=\left[M(s)\right]^*$, it leads to the following relation valid in the upper half plane \footnote{This  is identical to the case of massless branon scattering on the QCD flux tube discussed in \cite{FluxTube}.}
\beq
\la{crossingUHP}
 M(-s^*)=\left[M(s)\right]^*\,.
 \eeq


In addition, unitarity implies  positivity of the imaginary part of both forward scattering amplitudes.
More explicitly, we can set $t=0$ in equation \eqref{decomposition} to find 
\begin{align}
{\rm Im} \, M_{\pi_0\pi_0}(s) &= \frac{16 \pi}{3} \sum_l (2l+1)   \,{\rm Re}\left[ 3-S_l^{(0)}(s)-2S_l^{(2)}(s)\right]\ge 0\\
{\rm Im} \, M_t (s) &= 8 \pi \sum_l (2l+1)   \,{\rm Re}\left[ 2-S_l^{(1)}(s)-S_l^{(2)}(s)\right]\ge 0
\end{align}

Finally, using the results of appendix \ref{ap:PU} with $N=3$, we can write the low energy expansions
\begin{align}
M_{\pi_0\pi_0}(s)&=2(\alpha+2\beta)s^2 -\frac{1}{16\pi^2}s^2\left[
\log(s)+\log(-s)\right]+\dots
\label{Mlowenergy}\\
M_t(s)&=2 \beta s^2 -\frac{1}{48\pi^2}s^2\left[
\log(s)+\log(-s)\right]+\dots
\label{Mlowenergy2}
\end{align}
Here we use the standard convention of placing the cut of the logarithm along the negative real axis of its argument. Evaluating these expressions at $s+i\epsilon$ for positive real $s$, we find
\begin{align}
\IM M_{\pi_0\pi_0}(s)&=\frac{s^2}{16 \pi}+\frac{1}{13824\pi^3}(576\pi^2(3\alpha+2\beta)-1)s^3-\frac{5}{1152\pi^3}s^3\log(s)+\dots
\label{ImMlowenergy}\\
\IM M_t(s)&=\frac{s^2}{48 \pi}-\frac{3+64 \pi^2 (3\alpha+7\beta)}{6144\pi^3}s^3+\frac{5}{2304\pi^3}s^3\log{s}+\dots
\label{ImMlowenergy2}
\end{align}
where we included also the 2-loop contributions to the imaginary part.


\begin{figure}[t!]
  \centering
    \includegraphics[width=.7
     \textwidth]{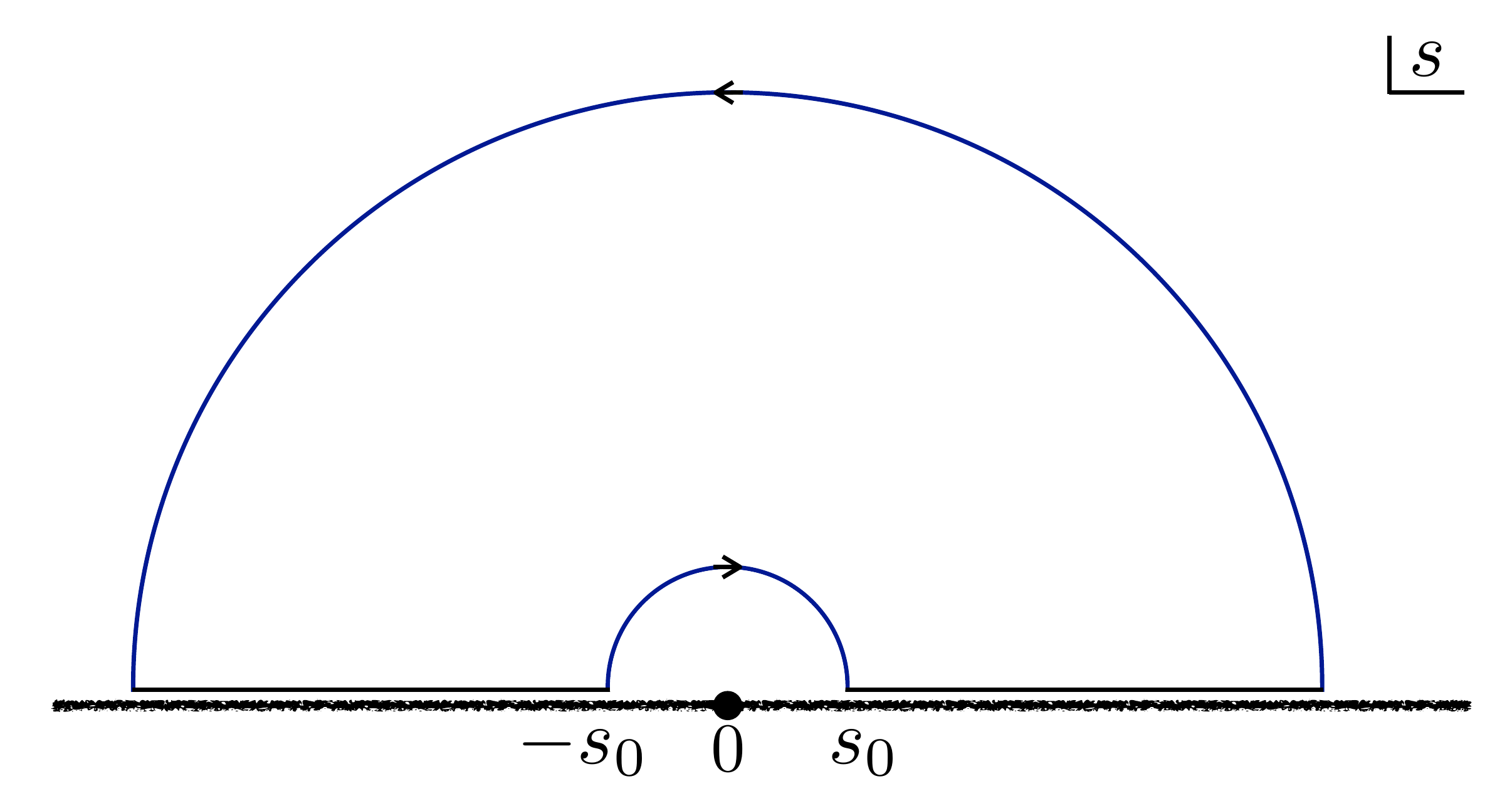}
  \caption{Closed integration contour in the $s$-complex plane used to derive  \eqref{arcint}.
  We assume the integrand decays sufficiently fast at large $|s|$ so that we can drop the arc at infinity.
  The forward amplitudes are analytic in the upper half plane and satisfy the crossing property \eqref{crossingUHP}. There are branch cuts along the real axis emanating from the branch point $s=0$.
   }\label{fig:arcs}
\end{figure}

\begin{figure}[t!]
  \centering
    \includegraphics[width=1
     \textwidth]{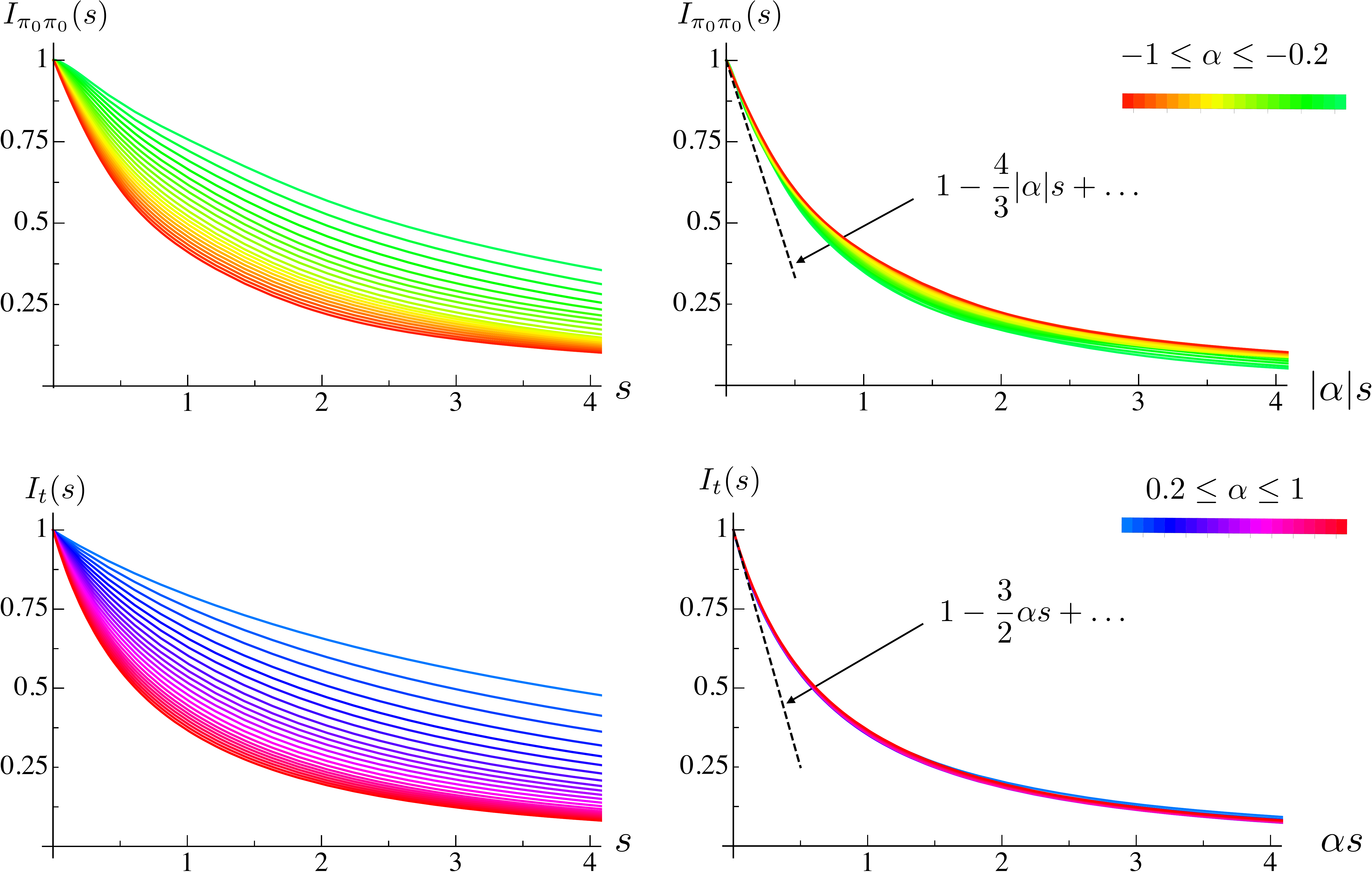}
  \caption{
  Imaginary parts of the two crossing symmetric forward amplitudes $M_{\pi^0\pi^0}(s)$ and $M_t(s)$.
  More precisely, on the left plots, we show the ratio $I(s)$ defined in \eqref{Iofs} for the numerical amplitude along the boundary of the allowed region shown in the inset of figure \ref{fig1} (for $N_\text{max}=23$).
  On the right, we test the scaling scenario \eqref{scalingscenario} by plotting the same data as a function of $|\alpha|s$. The curves collapse reasonably well, specially in the bottom plot.
  The black dashed lines are the analytic prediction from \eqref{ImMlowenergy} and \eqref{ImMlowenergy2}.
  Notice that this result follows from the 2-loop unitarity computation of appendix \ref{2loopAp}. It is not surprising that our numerics approximately match this prediction because unitarity is rather well saturated at low energies.
   }\label{fig:scaling}
\end{figure}

\subsection{Dispersive argument}
\la{arcsderivation}

The argument that follows applies to both forward amplitudes $M_{\pi_0\pi_0}(s)$ and $M_t(s)$. To avoid cluttering, we will denote the forward amplitude simply by $M(s)$. We shall use the properties of crossing symmetry  \eqref{crossingUHP}, positivity of the imaginary part ${\rm Im}\,M(s)\ge 0$ for $s>0$ and the low energy expansions
\beq
M(s) = c_0 s^2 - c_1s^2 \left[
\log(s)+\log(-s)\right] +\dots
\la{lowM}
\eeq
with the constants $c_0$ and $c_1$ given explicitly in \eqref{Mlowenergy} and \eqref{Mlowenergy2}.

%

Integrating $M(s)/s^3$ over the closed contour shown in figure \ref{fig:arcs}, we obtain the following relation
\beq
\frac{1}{s_0^2} \int_0^\pi d\theta e^{-2i\theta} M(s_0 e^{i\theta} ) 
= \int_{s_0}^\infty \frac{ds}{s^3} 2 {\rm Im} M(s)\,.
\la{arcint}
\eeq
We have made the usual assumption that $\lim_{|s|\to \infty} \frac{1}{s^2}{\rm Im} M(s) =0$.\footnote{For massive theories this follows from the Froissart bound \cite{Froissart:1961ux,Martin:1962rt}.}
Replacing the low energy expansion \eqref{lowM}, we find\footnote{Notice that we should replace $\log(s) =\log(s_0 e^{i\theta})$ and 
$\log(-s) =\log(s_0 e^{i(\theta-\pi)})$.}
\beq
 c_0 = 2 c_1 \lim_{s_0\to 0} \left[
 \int_{s_0}^\infty \frac{ds}{s} \frac{ {\rm Im} M(s) }{\pi c_1 s^2}+  \log s_0
 \right]\,.
 \la{sumrule}
\eeq
Notice that if we neglect the last logarithm, then one would immediately conclude that $c_0\ge 0$, which leads to the naive bounds \eqref{naivebounds}. This is essentially the argument of \cite{Pham:1985cr, Adams:2006sv}.
However, the last logarithm is crucial to obtain a finite limit when $s_0 \to 0$.
\footnote{Of course the authors of \cite{Adams:2006sv} were aware of the log branch cut in the forward amplitude. In fact, they describe its effect as logarithmic running of the couplings. Notice that we defined $\alpha$ and $\beta$ from the amplitude \eqref{lowenansatz}, where we fixed the scale of the logarithms  to $f_\pi$. The article \cite{Pham:1985cr} did not have to face this issue because it dealt with massive pions.}

It is convenient to define
\beq
I(s) \equiv \frac{ {\rm Im} M(s) }{\pi c_1 s^2} = 1 +O(s)\,.
\la{Iofs}
\eeq
in order to study the asymptotic behavior of \eqref{sumrule} for large $|\alpha|$.
The perturbative expansions \eqref{ImMlowenergy} and \eqref{ImMlowenergy2} suggest the following scaling at large $|\alpha|$,
\beq
I(s) =  I_0(|\alpha| s) + s \log (s) I_1(|\alpha| s) +\dots
\la{scalingscenario}
\eeq
where $I_0(0)=1$.  In figure \ref{fig:scaling}, we plot $I(s)$ for both forward amplitudes obtained from the numerical minimization of $\beta$ for large values of $|\alpha|$. This supports the  scaling scenario \eqref{scalingscenario}. 
This scenario predicts
\begin{align}
 c_0 &= 2c_1 \lim_{s_0\to 0} \left[
 \int_{s_0}^{\infty} \frac{ds}{s}  
 \left[ I_0(|\alpha| s) + s \log (s) I_1(|\alpha| s) +\dots \right]
 +  \log s_0
 \right]\\
 & =
  2 c_1 \lim_{s_0\to 0} \left[
 \int_{|\alpha| s_0}^{\infty} \frac{dz}{z}  
  I_0(z)  
 +  \log s_0
 \right]+ O\left(\frac{\log|\alpha|}{\alpha}\right)\\
 &  =- 2 c_1 \log|\alpha| + 
 2 c_1 \lim_{\epsilon \to 0} \left[
 \int_{\epsilon}^{\infty} \frac{dz}{z}  
  I_0(z)  
 +  \log \epsilon
 \right]+ O\left(\frac{\log|\alpha|}{\alpha}\right)
\end{align}
Applying this result to 
 $M_{\pi_0\pi_0}(s)$ for negative $\alpha$  and to $M_t(s)$ for positive $\alpha$, we obtain \eqref{improvednaivebounds}.
 In the inset of figure \ref{fig2}, we see that this prediction works very well for large positive $\alpha$ but not so well for large negative $\alpha$. This is probably related to the fact that the curves did not perfectly collapse in the top right plot of figure \ref{fig:scaling}. It is unclear if this would improve for larger values of $|\alpha|$ and with better numerical data, or if   the scenario \eqref{scalingscenario} is incorrect.

\bibliographystyle{JHEP}
\bibliography{4dmassless}{}

\end{document}